\documentclass{ws-p8-50x6-00m}

\newcommand{\ie}{{\em i.e.}}

\newcommand{\ccbar}{$c\bar{c}$}

\newcommand{\etal}{{\it et al.}}
\newcommand{\jpsi}{$J/\psi$}

\newcommand{\psip}{${\psi}^{\prime} \,$}

\def\ccb{$c\overline{c} \,\,$}

\def\mccb{m_{c\overline{c}}}
\def\EPJC{ {\em Eur. Phys. J.} C}
\newcommand{\lw}{\MakeLowercase}

\begin{document}

\title{Improved mapping of \ccbar \ onto charmonium states
\thanks {T\lw{o appear in the} P\lw{roceedings of} VIII I\lw{nternational} W\lw{orkshop on} H\lw{adron} P\lw{hysics} (Hadrons 2002), B\lw{ento} G\lw{on\c{c}alves}, B\lw{razil}, 14 - 19 A\lw{pril} 2002.}}
\author{C.~Brenner~Mariotto$^{1,2}$, M.B. Gay Ducati$^{1}$, G.~Ingelman$^{2,3}$}

\address{$^1$ Institute~of~Physics,~Univ.~Fed.~do~Rio~Grande~do~Sul,~Box~15051,~CEP~91501-960~Porto~Alegre,~Brazil\\ 
$^2$ High Energy Physics, Uppsala University, Box 535, S-75121 Uppsala, Sweden \\ 
$^3$ Deutsches~Elektronen-Synchrotron~DESY, Hamburg,~Germany\\
E-mail: mariotto@if.ufrgs.br}  

\maketitle

\abstracts{
We discuss the relative rates of different charmonium states and introduce an improved model for mapping the continuous \ccbar \ mass spectrum on the physical charmonium resonances. }

The theoretical description of charmonium production separates the hard and soft parts of the process based on the factorisation theorem in QCD. 
Perturbative QCD (pQCD) should be applicable for \ccbar \ production, since the charm quark mass $m_c$ is large enough to make $\alpha_s(m_c^2)$ a small expansion parameter. The formation of bound hadron states occurs through processes with small momentum transfers making $\alpha_s$ large and preventing the use of perturbation theory. This forces us to use phenomenological models to describe the formation of charmonium states from perturbatively produced \ccbar \ pairs. The Colour Evaporation (CEM) \cite{HALZENquantit} and Soft Colour Interaction (SCI) \cite{sci,sci-onium} models are based on a similar phenomenological approach, where soft colour interactions can change the colour state of a \ccbar \ pair from octet to singlet. They employ the same hard pQCD processes to produce a \ccbar \ pair regardless of its spin state. A colour singlet \ccbar \ pair with an invariant mass below the threshold for open charm ($m_{c\bar{c}}<2m_D$) will then form a charmonium state. Mapping \ccbar \ pairs onto charmonium states is based on phenomenological parameters taken from comparison to data (CEM), or from spin statistics (SCI), the latter resulting in a fraction of a specific quarkonium state $i$ with total angular momentum $J_i$ and main quantum number $n_i$ given by $f_i = \frac{\Gamma_i}{\sum_k \Gamma_k} $, where $\Gamma = (2J_i+1)/n_i$.
The CEM and SCI models give in general a good description of charmonium data both from fixed target and Tevatron energies \cite{sci-onium,cemmar,damet}. However, the overall energy-independent factors which describe the mapping of \ccb pairs onto different charmonium states are not satisfactorily understood. In particular, spin statistics gives an acceptable ratio $\psi^\prime /\psi$ at the Tevatron~\cite{sci-onium}, but fails at fixed target energies where an extra suppression by a factor four is required\cite{cemmar}. In other words, this ratio has an energy dependence which has not been accounted for. In this work we try to understand this problem by developing a model which introduces a correlation between the invariant mass of the \ccbar \ pair and the mass of the final charmonium state. 

The \ccbar \ pair is produced in a pQCD process with a continuous distribution of its invariant mass $m_{c\bar{c}}$ and must be mapped onto the discrete spectrum of charmonium states. The soft interactions that turn the pair into a colour singlet may very well change its mass by a few hundred MeV, which is the typical scale of the soft interactions, but larger mass shifts should be suppressed. The different charmonium states ($\eta_c$, \jpsi, $\chi_c$, \psip) are, however, separated in mass over a region of about 1~GeV. It therefore seems likely that the probability to form a particular charmonium state will depend on the original value of $m_{c\bar{c}}$ and not just on an overall factor, such as spin statistics. Thus, there should be a relatively larger probability to form states that are nearby in mass, than those further away on the mass scale. Based on these considerations we have constructed the following model. 

The smearing of the \ccbar \ mass due to soft interactions is considered as
\begin{equation}
G_{sme}({\mccb}, m)= \exp\left( -\frac{(\mccb-m)^2}{2\sigma_{sme}^2}\right) ,
\label{gauss1}
\end{equation}
where the gaussian width $\sigma_{sme}$ should be a few hundred MeV. 
The different charmonium resonances $i$ are represented by functions $F_{i}(m_i, m)$ around their physical masses $m_i$. For a \ccbar \ pair with a given mass $m_{c\bar{c}}$, the probability to form a certain charmonium state is then 
\begin{eqnarray}
{\cal {P}}_{i}({\mccb})&=& \frac {\int  G_{sme}({\mccb}, m) F_{i}(m_i, m) dm}
{\sum_{j} \int  G_{sme}({\mccb}, m) F_{j}(m_j, m) {dm}}
\approx 
\frac { s_i G_{sme}({\mccb}, m_i) }{\sum_{j} s_j G_{sme}({\mccb}, m_j) } , 
\label{Ps}
\end{eqnarray}
where the sum is over all charmonium resonances. The approximate result, which is used in our practical calculations, is obtained by letting $F_{i}(m_i, m)=s_i\delta (m-m_i)$, \ie \ neglecting the width of the very narrow charmonium states and including the previously used spin statistical factor  $s_i=(2J_i+1)/n_i$. 

\begin{figure}[t]
\begin{center}
  \includegraphics[height=.32\textheight]{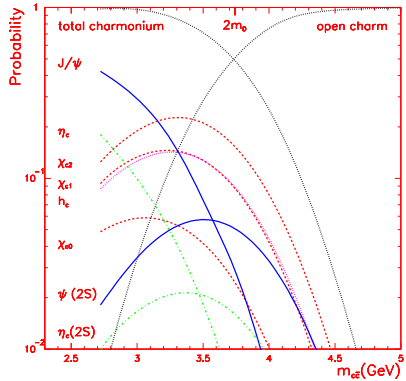}
  \caption{Probability distributions for the different charmonium states as obtained in the model with gaussian smearing ($\sigma_{sme}=400\, MeV$). The resulting total probability for charmonium production and the remainder as open charm production are also shown.}
\label{dQ2PsDserr}
\end{center}
\vspace{-0.5cm}
\end{figure}

The smearing of $m_{c\bar{c}}$ across the threshold $2m_D$ for open charm, implies non-zero contributions for charmonium also above the $D\overline{D}$ threshold as well as some open charm production for $m_{c\bar{c}}$ originally below this threshold.
 For a given original $m_{c\bar{c}}$ value, the probability to end up in the region above $D\overline{D}$ threshold is given by the area of the smearing gaussian in that region, \ie \ 
\begin{eqnarray}
A(m_{c\overline{c}})&=& \frac {1}{\sqrt{2\pi}\sigma_{sme}} \int_{2m_D}^{\infty} \exp\left( -\frac{(\mccb-m)^2}{2\sigma_{sme}^2}\right) dm
\,.
\end{eqnarray}
The probability to end up in a specific resonance is then given by $(1-A){\cal{P}}_i$ and is shown in Fig.~\ref{dQ2PsDserr} for the different charmonium states. For a given original \ccbar \ mass $m_{c\bar{c}}$ one can here see the fractional production of different states, as well as the sum of all charmonium states together with the remainder giving open charm.

By folding these charmonium probability functions with the distribution in $m_{c\bar{c}}$ obtained from pQCD, \ie \ ${d\sigma}_{c\overline{c}} /dm_{c\overline{c}}$ (LO processes and $m_c=1.35$\ GeV) , the cross section for a given charmonium state is obtained as 
\begin{equation}
\sigma_{i}= \int_{2m_c}^{\sqrt{s}} dm_{c\overline{c}} 
\frac{{d\sigma}_{c\overline{c}}} {dm_{c\overline{c}}}(1-A(\mccb)){\cal {P}}_{i}({\mccb})\, .
\end{equation}

\begin{figure}[t]
\begin{tabular}{c c}
  \includegraphics[height=.29\textheight]{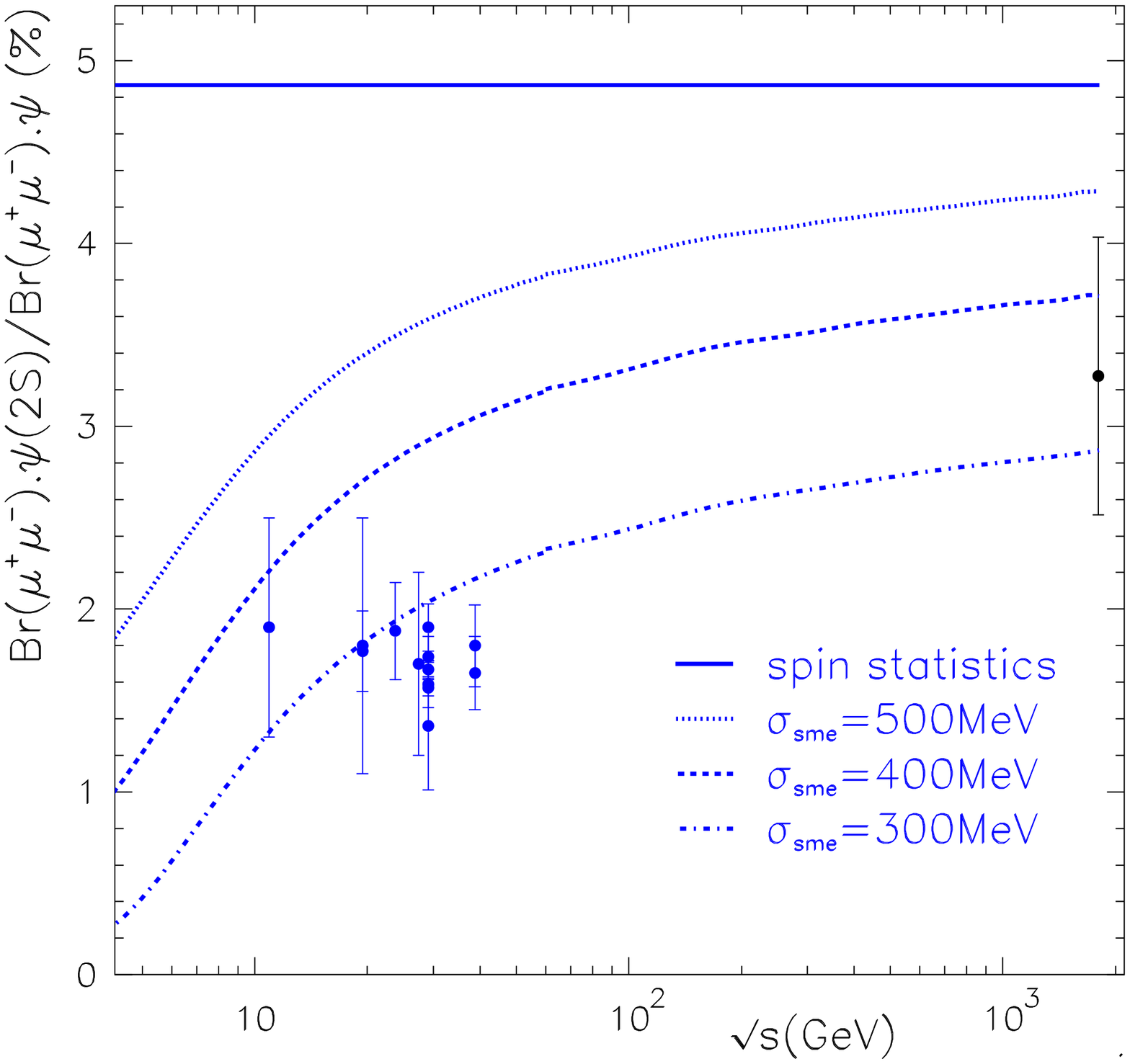}
  \includegraphics[height=.40\textheight]{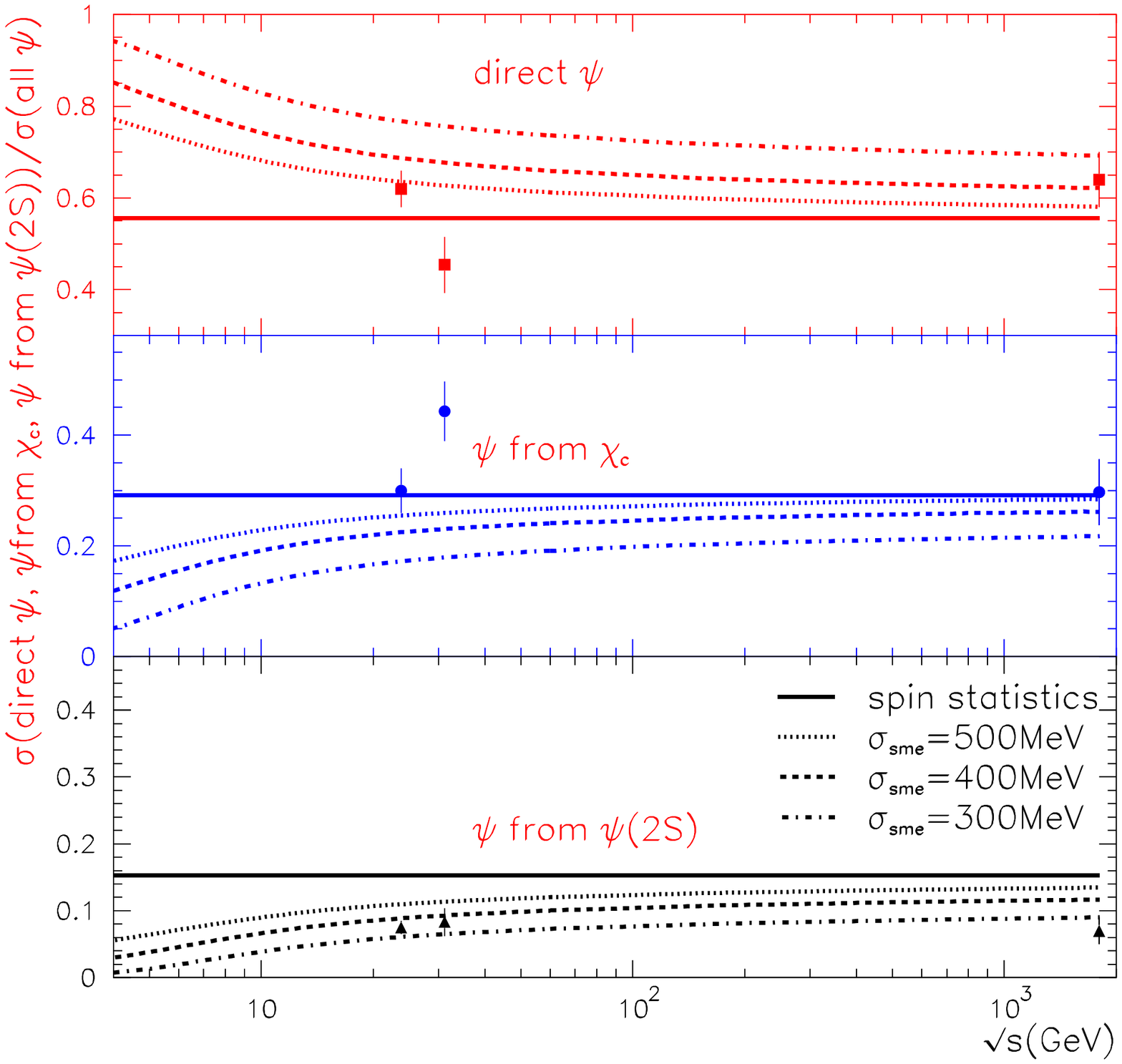}
\end{tabular}
\caption{The ratio \psip / \jpsi \ (times their branching ratios for decay into $\mu^+\mu^-$) (left) and fractions of \jpsi \ produced directly, and coming from the decay of $\chi_c$ and $\psi^{\prime}$ states (right) in hadron-hadron interactions of cms energy $\sqrt{s}$. Data ${\tiny {}^{6,7,8}} $ compared to simple spin statistics and to our model with different gaussian smearing widths applied to CEM.}
\label{dQ2gaussfract}
\vspace{-0.4cm}
\end{figure}

Applying this mapping procedure to the CEM model we obtain the results in Fig.~\ref{dQ2gaussfract}. As opposed to the simple spin statistics factor, this model gives a reasonable description of the observed ratio of \psip \ to \jpsi \ production and fractions of \jpsi \ produced directly, coming from decays of $\chi_c$ states and from $\psi^{\prime}$. In particular, the model gives a characteristic energy dependence of the kind indicated by the data. 
Considering all observables from Fig.~\ref{dQ2gaussfract}, the preferred value for the $\sigma_{sme}$ is 400~MeV or lower. This fits well with the expectation for the non-perturbative dynamics that this model should describe. The essential new ingredient of this model, namely introducing a correlation between the invariant mass of the \ccbar \ pair and the masses of the different charmonium states, is found to give an improved description of the ratios of different charmonium states as compared to using constant factors based on spin statistics. In particular, these ratios acquire an energy dependence which is particularly strong at low energies where threshold effects are more pronounced.

{\bf Acknowledgments:}  This work was partially financed by CAPES, Brazil, and by the Swedish Research Council.

\end{document}